\documentclass{article}
\usepackage{waspaa23}
\usepackage{amssymb,amsmath,graphicx,times,url,bm}
\usepackage{color,cite}
\usepackage{booktabs,tabularx}
\usepackage{comment}
\usepackage{hyperref}
\usepackage{multirow}
\usepackage{tipa,tipx}
\usepackage{balance} 
\usepackage{footmisc}

\newcommand{\demourl}{\url{google.github.io/df-conformer/miipher/}}

\DeclareMathOperator{\film}{FiLM}
\DeclareMathOperator{\lrelu}{LeakyReLU}
\DeclareMathOperator{\CNN}{CNN}
\DeclareMathOperator{\abs}{abs}

\usepackage{ifthen}
\title{Miipher: A Robust Speech Restoration Model Integrating \\ Self-Supervised Speech and Text Representations}
\name{
\parbox{0.99\linewidth}{
\centering
Yuma~Koizumi$^1$,
Heiga~Zen$^1$,
Shigeki~Karita$^1$,
Yifan~Ding$^1$,
Kohei~Yatabe$^2$,
Nobuyuki~Morioka$^1$,\\
Yu~Zhang$^3$,
Wei~Han$^3$,
Ankur~Bapna$^3$,
Michiel~Bacchiani$^1$
}}
\address{\hspace{-14pt}$^1\,$Google, Japan \hspace{18pt} $^2\,$Tokyo University of Agriculture \& Technology, Japan \hspace{18pt} $^3\,$Google, USA}
\begin{document}

\ninept
\maketitle

\begin{sloppy}

\begin{abstract}
Speech restoration (SR) is a task of converting degraded speech signals into high-quality ones. In this study, we propose a robust SR model called \textit{Miipher}, and apply Miipher to a new SR application: increasing the amount of high-quality training data for speech generation by converting speech samples collected from the Web to studio-quality. To make our SR model robust against various degradation, we use (i) a speech representation extracted from w2v-BERT for the input feature, and (ii) a text representation extracted from transcripts via PnG-BERT as a linguistic conditioning feature. Experiments show that Miipher (i) is robust against various audio degradation and (ii) enable us to train a high-quality text-to-speech (TTS) model from restored speech samples collected from the Web.
Audio samples are available at our demo page: \demourl.
\end{abstract}


\section{Introduction}
\label{sec:intro}

Speech restoration (SR) is a task of converting degraded speech signals into high-quality speech signals~\cite{Maiti_waspaa_2019,self_remaster,Su_2021,voice_filxer,UNIVERSE}. It is a comprehensive task including enhancement~\cite{dlwang_2018}, 
dereverberation~\cite{wpe,han_2014,dnn_wpe},
%dereverberation~\cite{wpe},
declipping~\cite{declip_survey}, and
super-resolution~\cite{Li_2015,Kuleshov_2017}.
%super-resolution~\cite{Li_2015}.
SR has been addressed to improve speech intelligibility for helping human listening, and its quality has evolved over the past few years with the evolution of the speech generative models~\cite{wavenet,tacotron2,parrotron}.
Parametric resynthesis~\cite{Maiti_waspaa_2019}-based methods~\cite{self_remaster,Su_2021,voice_filxer} and diffusion model-based methods~\cite{UNIVERSE} have been proposed.
They can convert lecture and historical speech into high-quality speech as if these were recorded in a studio.

By converting speech samples in-the-wild to studio-quality using advanced SR methods, we are trying to increase the amount of high-quality training data for speech generation.
The performance of deep learning depends on both size and quality of the training dataset~\cite{speechstew,bigvgan,wavefit}.
As the cost for collecting studio-recorded samples is expensive, recent research has tried to alleviate the requirement for dataset by developing specific schemes, such as self-supervised learning (SSL)~\cite{w2v22020,w2vbert,bestrq}, and using (low-quality) speech samples from the Web~\cite{common_voice,vox_populi}.
However, unlike automatic speech recognition (ASR), training of text-to-speech (TTS) models using samples in-the-wild is still a challenging
problem~\cite{hsu2018hierarchical,hsu_2019}
%problem~\cite{hsu_2019}
because the quality of speech generation is directly affected by that of the training samples.

The main challenge for the application of SR in this study is the robustness aspect.
If SR fails to restore speech samples and produces artifacts, they will negatively affect the subsequent training
 of speech generative models.
Thus, we must reduce the number of restoration failed samples as much as possible.

In this study, we propose \textit{Miipher}\footnote{Miipher:  \underline{\textbf{m}}ult\underline{\textbf{i}}ple features \underline{\textbf{i}}ntegrated s\underline{\textbf{p}}eec\underline{\textbf{h}} r\underline{\textbf{e}}sto\underline{\textbf{r}}ation}, a robust parametric re-synthesis SR model for restoring speech samples in-the-wild.
Based on our preliminary investigation, we especially focus on the following two difficult degradations where SR frequently fails:
\begin{description}
 \setlength{\itemsep}{0pt}
 \item{\textbf{Phoneme masking}}: Speech signals are sometimes masked by noise and/or reverberation.
 It is difficult to discriminate speech and noise without having additional information.
 \item{\textbf{Phoneme deletion}}: Important frequency components of some phonemes could be missing from the signal due to codecs and/or down-sampling. 
 If a noisy sample lacks a phoneme, it introduces an unrecoverable error.
\end{description}
To solve these issues, we introduce the following two techniques:
\begin{description}
 \setlength{\itemsep}{0pt}
 \item{\textbf{SSL features domain cleaning}}:
 For the input feature, instead of a log-mel spectrogram used in conventional methods~\cite{Maiti_waspaa_2019,voice_filxer}, we use a speech representation extracted from w2v-BERT~\cite{w2vbert}, an SSL model trained on degraded speech samples.
 As it improves ASR performance, we expect its effectiveness on making SR models robust against speech degradation.
 \item{\textbf{Transcript conditioning}}: We consider the deleted phoneme reconstruction problem as a text-conditioned speech inpainting~\cite{borsos22_interspeech}.
 As a linguistic conditioning feature, we use a text representation extracted from PnG-BERT~\cite{pngbert}, a text SSL model for TTS use.
\end{description}
Through experiments, we show that 
(i) Miipher is robust against various audio degradations, and
(ii) applying Miipher on a large scale speech dataset collected from the Web enables us to train a TTS model whose quality is on par with one that is trained on a dataset designed for TTS training.
Audio samples of the restored and TTS generated samples are available at our demo page\footnote{\demourl\label{footnote:demo}}.

\section{Speech restoration model}
\label{sec:model}

\subsection{Model overview}
\label{sec:overview}

\begin{figure}[t]
% \vspace{-1pt}
  \centering
  \includegraphics[width=\linewidth,clip]{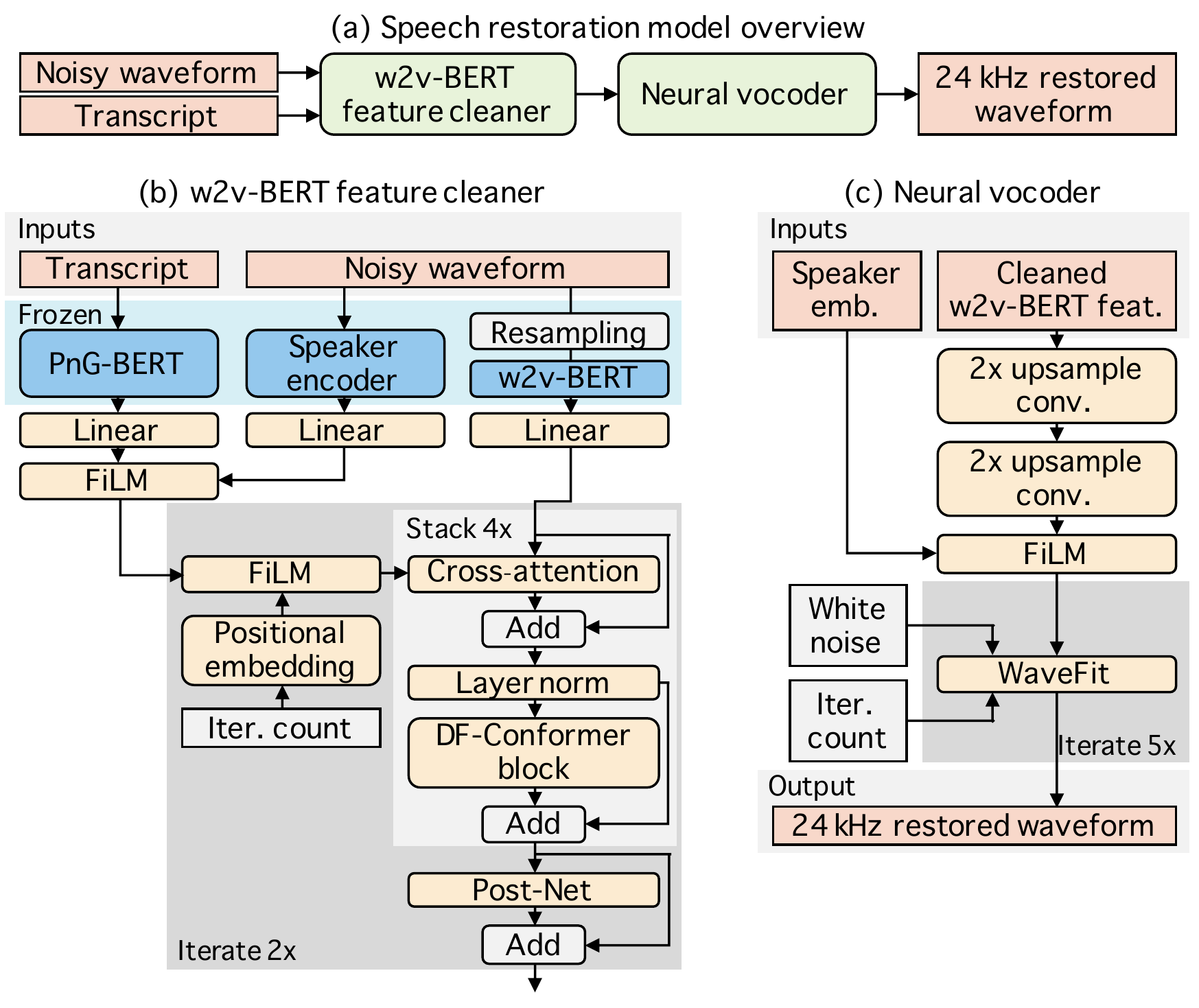} 
  \vspace{-22pt}
  \caption{(a) Overview of Miipher. (b) w2v-BERT and PnG-BERT are used as feature extractors, and a DF-Conformer-based feature cleaner predicts clean w2v-BERT features. Then, (c) a WaveFit synthesizes a restored waveform from cleaned w2v-BERT features.}
  \label{fig:model}
\end{figure}

Let the $T$-sample time-domain signal $\bm{x} \in \mathbb{R}^T$ be a degraded signal of an original signal $\bm{s} \in \mathbb{R}^{T}$.
The goal of SR is to recover $\bm{s}$ from $\bm{x}$ as $\bm{y} \in \mathbb{R}^{T}$. In this study, the sampling rate of $\bm{s}$ and $\bm{y}$ is 24\,kHz, because 24\,kHz sampling signals are often used in speech generation tasks such as TTS~\cite{tacotron2}.

To achieve this goal, we use a parametric re-synthesis framework~\cite{Maiti_waspaa_2019} as shown in Fig.~\ref{fig:model}~(a): a noisy feature is cleaned by a feature cleaner network, then the restored wavefrom is synthesized using a neural vocoder.
To make our SR model robust against various speech degradation, we used a speech representation extracted from w2v-BERT~\cite{w2vbert} for the input feature of SR, and linguistic features conditioning extracted by PnG-BERT~\cite{pngbert} from the transcript.

\subsection{Feature cleaner}
\label{sec:feature_enhancer}
The overview of the feature cleaner is shown in Fig.~\ref{fig:model}~(b). Our feature cleaner predicts w2v-BERT features of the clean speech 
$\bm{S} \in \mathbb{R}^{K \times D}$ from that of degraded speech $\bm{X} \in \mathbb{R}^{K \times D}$ as $\hat{\bm{S}} \in \mathbb{R}^{K \times D}$.
Here, $K$ is the number of time-frames and $D$ is the dimension of w2v-BERT feature.
As a conditioning of SR, we use linguistic features extracted from a transcript via PnG-BERT $\bm{E} \in \mathbb{R}^{M \times W}$ where $M$ is the number of tokens and $W$ is the dimension of PnG-BERT features.
In addition, since SSL features often lose speaker information~\cite{hubert_neural_vocoder}, we use a speaker embedding $\bm{d} \in \mathbb{R}^{Q}$ extracted from $\bm{x}$ as speaker conditioning.
Note that since w2v-BERT is trained from 16\,kHz audio, we resample $\bm{x}$ at 16\,kHz for extracting $\bm{X}$.

First, the feature dimension of $\bm{X}$, $\bm{d}$ and $\bm{E}$ are aligned to $D_b$ by linear layers, where $D_b$ is the input dimension of the DF-Conformer block.
Then, $\bm{d}$ is combined to $\bm{E}$ using a CNN-based simple feature-wise linear modulation (FiLM) layer~\cite{film}.
Here a FiLM layer is defined as
$\film(\bm{A}, \bm{b}) = \CNN_2 \left( \lrelu(\CNN_1(\bm{A})) + \bm{b} \right)$
where $\bm{A} \in \mathbb{R}^{K \times D}$ and $\bm{b} \in \mathbb{R}^{Q}$.
$\CNN_1$ changes $\bm{A}$'s 2nd dimension $D$ to $Q$. Then, $\bm{b}$ is added to the $\CNN_1$ output while broadcasting the 1st dimension. Here, the slope parameter of $\lrelu$ is 0.1. Finally, $\CNN_2$ returns the 2nd dimension to $D$. Both CNNs are 1D-CNN layer with a kernel-size 3 and stride 1.

Next, $\bm{S}$ is predicted from $\bm{X}$ by using a DF-Conformer~\cite{Koizumi_waspaa_2021}-based network.
We build our feature cleaner by stacking $N$ blocks, each consisting of a cross attention, a layer norm and a DF-Conformer-block.
In each block, the combined conditioning feature is combined to the input w2v-BERT features using a cross attention layer. Then, layer norm is applied to the output, and finally, a DF-Conformer block~\cite{Koizumi_waspaa_2021} with $\mod(n, 2)$ dilation factor is applied where $n \in \{0,1,...,N-1\}$ is the block number.
After refining the input w2v-BERT features, we applied the 5-layer convolutional Post-Net proposed in Tacotron2~\cite{tacotron2} which predicts a residual to add to the prediction to improve the overall reconstruction.

We adopt the iterative feature refinement into our feature cleaner since a fixed-point iteration-like processing achieved high-fidelity waveform generation~\cite{wavefit}. Specifically, we iterate twice the entire feature cleaning process consisting of the feature cleaner and the Post-Net, where the parameters of the layers are shared.
As with WaveGrad~\cite{wavegrad} and WaveFit~\cite{wavefit}, the number of iterations is embedded using a positional embedding layer, and
the embedded iteration count is mixed into text and speaker embedding using a FiLM layer.
Note that it is not strictly fixed-point iteration since the function being applied is slightly different at each iteration.

%As the loss function, we used a combination 
We used a combined loss function
of the mean-absolute-error, the mean-squared-error, and a spectral convergence loss~\cite{parallel_wavegan}-like loss as
$\mathcal{L} = \lVert \bm{S} - \hat{\bm{S}} \rVert_{1} + \lVert \bm{S} - \hat{\bm{S}} \rVert_{2}^2 + \lVert \bm{S} - \hat{\bm{S}} \rVert_{2}^2 / \lVert \bm{S} \rVert_{2}^2,$
where $\lVert \cdot \rVert_p$ is the vector $p$-norm given by $\lVert \bm{S} \rVert _p = ( \sum_k \sum_d |S_{k,d}|^p )^{1/p}$.
This loss value is calculated before and after the Post-Net like in Tacotron2~\cite{tacotron2}, and calculated for all iterations like in WaveFit~\cite{wavefit}.

\subsection{Neural vocoder}
\label{sec:wavefit}

As shown in Fig.~\ref{fig:model}~(c), we use a WaveFit neural vocoder~\cite{wavefit} to synthesize $\bm{y}$ from $\hat{\bm{S}} \in \mathbb{R}^{K \times D}$.
The original WaveFit synthesizes waveforms from log-Mel spectrograms; here we adopt WaveFit to synthesize waveforms from w2v-BERT features.

First, the frame rate of w2v-BERT feature and log-mel spectrogram does not match, hence $\hat{\bm{S}}$ is 4x upsampled along the time axis.
Since the network architecture of WaveFit consists of a CNN-based U-Net~\cite{wavegrad}, a deep upsampling network is required to synthesize a waveform from low frame-rate features.
To keep the core network size close to the original one, we applied two
upsampling CNN layers: each layer consists of a transposed convolution with ReLU, channel size 1 where the kernel- and stride-sizes are 4 and 2, respectively.
Then, the conditioning speaker embedding $\bm{d}$ is combined with the upsampled w2v-BERT feature $\hat{\bm{S}}' \in \mathbb{R}^{4K \times D}$ using a FiLM layer as $\film(\hat{\bm{S}}', \bm{d})$.

Finally, the speech waveform is synthesized by WaveFit. Note that we used white noise as the initial noise waveform instead of the SpecGrad noise~\cite{specgrad} used in the original WaveFit~\cite{wavefit}
because we cannot know the spectral envelope of $\bm{s}$ from w2v-BERT features.
In addition, since the maximum amplitude of $\bm{s}$ cannot be known from w2v-BERT features, the gain normalization of WaveFit is replaced to $\mathcal{G}(\bm{y}) = \lambda \cdot \bm{y} / \max(\abs(\bm{y}))$  where $\lambda = 0.9$ and $\abs(\cdot)$ returns element-wise absolute values of the input vector.

For the loss function, in addition to the original adversarial loss function proposed in WaveFit~\cite{wavefit}, we used the multi-period discriminator (MPD)~\cite{Kong_2020}.
We used the same MPD hyper-parameters used in \cite{Kong_2020} except for additionally using three more prime-number periods (13, 17, 19) because the sampling rate of $\bm{y}$ is 24\,kHz.
In addition, the parameters of the STFT loss was changed to the same as those of Parallel WaveGAN settings~\cite{parallel_wavegan}.

\subsection{Model training}
\label{sec:training}

\textbf{Frozen feature extraction models:}
We used the ``w2v-BERT XL'' model, where $D=1024$, trained on 60k hours of English speech samples from the LibriVox repository.
We used the 8th layer Conformer outputs without quantization as w2v-BERT features.
For the PnG-BERT model, we used a $W=512$ PnG-BERT model~\cite{pngbert} pre-trained on a plain text corpus mined from Wikipedia, containing 131M English sentences.
To extract speaker embedding, we used a streaming Conformer-based speaker encoding model consists of 12 Conformer~\cite{conformer} layers each of $Q=256$, followed by an attentive temporal pooling module~\cite{quan2022_odyssey}.
The model was trained on the dataset described in \cite{dvector_data} while minimizing the generalized end-to-end extended-set softmax (GE2E-XS) loss~\cite{drvector}.

\vspace{2pt}
\noindent
\textbf{Trainable model parameters:} 
We stacked $N=4$ DF-Conformer blocks~\cite{Koizumi_waspaa_2021} where the input and hidden dimensions of the multi-head attention are $D_b = 128$ and 512, respectively.
Other hyper-parameters were the same as those of the original paper~\cite{Koizumi_waspaa_2021}.
For the cross-attention layer before the DF-Conformer block, the input and hidden dimensions were 128 and 512, respectively. The hyper-parameters of the Post-Net are the same as those of Tacotron2~\cite{tacotron2}. 
For the neural vocoder, we used WaveFit-5 with the same hyper-parameters described in \cite{wavefit} except for the upsampling factors of the five upsampling blocks (UBlock) are (5, 4, 2, 2, 2) because the frame rate of the upsampled w2v-BERT feature is 100 [frame/sec].
To accelerate training, we randomly picked up 15 frames (0.6 seconds) from $\bm{X}$ as input.

\begin{table}[ttt]
\caption{Random codec parameters. Codec was randomly selected based on the probability listed in the ``Prob.'' column. Codec bitrate was randomly selected from those listed in the ``Bit rate'' column.}
\label{tab:codec}
\centering
\begin{tabular}{l | c | c }
\toprule
\textbf{Codec} & \textbf{Prob.} & \textbf{Bit rate} \\
\midrule
MP3 & 0.5 & 16k, 32k, 64k, 128k \\
Vorbis & 0.075 & 32k, 48k, 64k \\
A-law & 0.025 & 64k \\
\multirow{2}{*}{AMR-WB} & \multirow{2}{*}{0.025} & 6.6k, 8.85k, 12.65k, 14.25k, 15.85k  \\
 &  & 18.25k, 19.85k, 23.05k, 23.85k \\
OPUS & 0.375 & 8k, 16k, 32k, 64k, 128k \\
\bottomrule
\end{tabular}
\end{table}

\vspace{2pt}
\noindent
\textbf{Training dataset:} We trained the proposed model with a proprietary dataset that contains 2,680 hours of noisy and studio-quality speech pairs.
The target speech dataset contains 670 hours of studio-recorded Australia, United Kingdom, India, Nigeria, and United States English at 24\,kHz sampling. 
For the noise dataset, we used the TAU Urban Audio-Visual Scenes 2021 dataset~\cite{tau_2021_dataset}, internally collected noise snippets that simulate conditions like cafe, kitchen, and cars, and noise and music sources.
The noisy utterances were generated by mixing randomly selected speech and noise samples from these datasets with signal-to-noise ratio (SNR) from 5\,dB to 30\,dB.
In addition, we augmented the noisy dataset with 4 patterns depending on the presence or absence of reverberation and codec artifacts.
A room impulse response (RIR) for each sample was generated by a stochastic RIR generator using the image method~\cite{image_method}. 
Its parameters were drawn from the following uniform distributions $\mathcal{U}$: the reverberation times (RT60) ranging from $\mathcal{U}(0.2, 0.5)$ [sec], the length of one side for x- and y-axis (width) room was $\mathcal{U}(2.0, 10.0)$ [m], and 
that for z-axis (height) $\mathcal{U}(2.0, 5.0)$ [m].
For simulating codec artifacts, we randomly applied one of MP3, Vorbis, A-law, Adaptive Multi-Rate Wideband (AMR-WB), and OPUS with a random bit-rate.
The codec parameters were shown in Table \ref{tab:codec} which were decided with reference to \cite{UNIVERSE}.

\section{Experiments}
\label{sec:experiment}

\subsection{Experimental setup}
\label{sec:steup}

We evaluated the effectiveness of w2v-BERT and PnG-BERT while comparing the sound quality between the studio-recorded original speech and restored samples from artificially contaminated samples.

\vspace{2pt}
\textbf{Test dataset:} Test dataset was generated by contaminating the clean data with the same way for generating the training dataset described in Sec.~\ref{sec:training}.
The clean speech samples were studio-recorded US-English containing 534 utterances spoken by 1 female and 1 male. There was no overlap of this data with the training set in terms of the utterances themselves or the speakers. The additive noise datset was the DEMAND noise dataset~\cite{demand}.
We applied reverberation to all clean speech samples using the same stochastic RIR generator with the same parameters as the training dataset.
After that, the noisy samples were generated by mixing randomly selected reverberant speech and noise samples from these datasets with SNR from 5\,dB to 30\,dB.
Finally, we simulated codec artifacts to the noisy samples with the same parameters with the training dataset.

\vspace{2pt}
\noindent
\textbf{Comparison methods:} The main novelties of the proposed method are to use (i) w2v-BERT features instead of log-mel spectrogram, (ii) PnG-BERT for text conditioning, and (iii) a speaker embedding for speaker conditioning. Therefore, we compared eight models with and without these three patterns: 
i.e. \{w2v-BERT, log-mel\} $\times$ \{full, w/o PnG-BERT, w/o speaker embedding, w/o PnG-BERT \& speaker embedding\}.
For log-mel models, we used 128-dimensional log-mel spectrograms (50\,ms Hann window, 12.5\,ms frame shift, 2048-point FFT, and 20\,Hz and 12\,kHz lower and upper frequency cutoffs, respectively).
As the neural vocoder for the log-mel models, we used the original WaveFit-5~\cite{wavefit} except for using the MPD~\cite{Kong_2020} with the same parameters as w2v-BERT models. 
Other parameters were the same among the all comparison models.
We trained the feature-cleaner 400k steps. Individually, we pre-trained the WaveFit 200k steps to reconstruct waveform from clean features, after that, we fine-tuned the WaveFit 100k steps to synthesize the clean waveform from the predicted features by the trained feature cleaner.

\subsection{Results}
\label{sec:result}

\begin{table}[ttt]
\caption{Results on the synthesized test dataset with their 95\% confidence intervals.
``L'', ``W''
``T'' and ``S'' means 
log-mel, w2v-BERT, transcript and speaker embedding conditioning, respectively.}
\label{tab:comparison_result}
\centering
\begin{tabular}{c l| c c c}
\toprule
\textbf{Feat.} & \textbf{Method}& \textbf{MOS} $(\uparrow)$ & \textbf{WER} $(\downarrow)$ & \textbf{SPK} $(\uparrow)$ \\	
\midrule
 - & Clean & $4.69 \pm 0.06$ & $13.7$ & N/A \\
 - & Noisy & $3.28 \pm 0.13$ & $15.1$ & $0.703$ \\
\midrule
\multirow{4}{*}{L} & Full             & $3.08 \pm 0.12$ & $19.4$ & $\bm{0.767}$ \\
        & w/o T      & $2.16 \pm 0.13$ & $39.6$ & $0.698$ \\
        & w/o S      & $3.17 \pm 0.08$ & $19.5$ & $0.711$ \\
        & w/o T \& S & $1.40 \pm 0.09$ & $61.4$ & $0.496$ \\
\midrule
\multirow{4}{*}{W} & Full (ours)      & $\bm{4.54} \pm \bm{0.08}$ & $\bm{13.5}$ & $0.727$ \\
        & w/o T      & $4.48 \pm 0.08$ & $16.4$ & $0.735$ \\
        & w/o S      & $4.39 \pm 0.08$ & $14.0$ & $0.646$ \\
        & w/o T \& S & $4.26 \pm 0.09$ & $17.2$ & $0.637$ \\
\bottomrule
\end{tabular}
\end{table}

To evaluate subjective quality, we rated speech quality through mean-opinion-score (MOS).
The scale of MOS was a 5-point scale (1:~Bad, 2:~Poor, 3:~Fair, 4:~Good, 5:~Excellent) with rating increments of 0.5.
Test stimuli were randomly chosen and each stimulus was evaluated by one subject. Each subject was allowed to evaluate up to six stimuli, that is, 89 subjects participated in this experiment to evaluate 534 samples in each condition. The subjects were paid native English speakers in the United States. They were requested to use headphones in a quiet room.
In addition, in order to confirm whether the text-content and speaker identity in the restored speech samples are retained, we evaluated the word error rate (WER) of an ASR experiment and computed a speaker similarity measure.
The WER was computed between the reference transcripts and the output of the ``Pre-trained Conformer XXL'' ASR model~\cite{yu_asr_2020}. 
Note that the ASR model was trained on noisy speech and transcripts normalized by a different text-normalizer. Therefore, we only check trends in WER differences since absolute value of WER for our evaluation dataset will be high.
To evaluate speaker similarity, we computed a cosine similarity (SPK) of the speaker embedding between the clean and restored samples~\cite{NEURIPS2018_6832a7b2,chen2018sample}.

Table~\ref{tab:comparison_result} shows the results.
The proposed method (w2v-BERT, full) achieved the best MOS among all SR comparison models. In addition, the MOS of the proposed model was only $0.15$ points behind that of the studio-recorded clean signals.
By comparing MOS of w2v-BERT models and log-mel models, the scores of w2v-BERT models were significantly higher. This result indicates that the use of w2v-BERT features has a significant effect on improving SR performance.
The WER of ``W w/o T'' was 2.9 point behind of the proposed method, and the WER of the proposed method was comparable to that of the clean signal. This shows that the use of PnG-BERT features extracted from a transcript is effective in preserving text content of the restored speech.
The SPK score of ``W w/o S'' model was significantly lower than that of the proposed method. This result indicates the use of a speaker embedding is effective for keeping speaker characteristics.
Audio samples are available at our demo page\footref{footnote:demo}.

\section{TTS model training with non-TTS dataset}
\label{sec:applications}

\begin{table}[ttt]
\caption{MOS scores with their 95\% confidence intervals of TTS outputs trained on each restored dataset.}
\label{tab:cv_vox_result}
\centering
\begin{tabular}{l | c c  }
\toprule
\textbf{Training 
dataset}  & \textbf{Speaker ID} & \textbf{MOS} ($\uparrow$) \\
\midrule
Common\,Voice & - & $3.63 \pm 0.21$ \\
\midrule
\multirow{4}{*}{LibriVox} & 4788 & $4.38 \pm 0.15$ \\
 & 10801 & $4.15 \pm 0.17$ \\
 & 5968& $4.11 \pm 0.19$ \\
 & 8138& $4.05 \pm 0.18$ \\
\bottomrule
\end{tabular}
\end{table}

Thanks to crowd-sourcing projects for creating license-free speech corpus, a huge amount of English text-speech pair data is publicly available. However, since these samples are not studio-recorded quality, their use-cases are mainly limited to ASR. 
To increase the value of such public speech samples, we restored the samples in Common\,Voice~\cite{common_voice} and LibriVox datasets using Miipher.
Then, we trained TTS models using the restored datasets. The details of this experiment is as follows.

\vspace{2pt}
\noindent
\textbf{TTS model:}
The TTS model consisted of an unsupervised Non-Attentive Tacotron (UNAT) with a fine-grained variational auto-encoder~\cite{nat} and a WaveRNN neural vocoder\cite{wavernn}.
For UNAT, we used the same hyper-parameters and training parameters listed in the original paper~\cite{nat}.
The WaveRNN~\cite{wavernn} consisted of a single long short-term memory layer with 512 hidden units, 5 convolutional layers with 512 channels as the conditioning stack to process the mel-spectrogram features, and a 10-component mixture of logistic distributions as its output layer. We trained UNAT and WaveRNN for 100k and 400k steps and concatenated them without fine-tuning.

\vspace{2pt}
\noindent
\textbf{Common\,Voice:} We restored samples in the version 5.1 (June 22, 2020) snapshot with approximately 1,500 hours. We removed speech samples of less than 2 seconds or more than 15 seconds. The training dataset consisted of 685 hours speech samples. Note that since Common\,Voice is prohibited from identifying the speaker, the TTS model was trained without using speaker IDs and trained as a single-speaker TTS model.

\vspace{2pt}
\noindent
\textbf{LibriVox:} We collected over 25,000 hours of speech samples (Jan.~7,~2020). There was no overlap of this data with LibriTTS~\cite{libritts}. We aligned the long-form audio recordings with the corresponding texts, and split them into sentence-level segments as the same way of building LibriTTS~\cite{libritts}. We removed speech samples of less than 2 seconds or more than 15 seconds. The training dataset consisted of 13,270 hours speech samples spoken by 4,000 speakers. We trained a multi-speaker TTS model.

We synthesized waveforms with the sentences from the LJspeech~\cite{ljspeech17} test split. For the LibriVox multi-speaker model, we used 4 speakers (two female and two male) where the female and male reader IDs were (14788, 10801) and (5968, 8168), respectively.

We evaluated the subjective sound quality using MOS.
To give baseline MOS on LJspeech test split sentences, we also trained the same TTS model using Miipher applied LJspeech~\cite{ljspeech17}.
The MOS of human spoken and TTS generated speech samples of LJspeech were 4.33 and 4.36 respectively

Table~\ref{tab:cv_vox_result} shows the MOS of each model.
Even though the LibriVox are not datasets designed for TTS like LJspeech, and the training of multi-speaker TTS is difficult compared with single-speaker TTS, TTS models trained using restored LibriVox achieved MOS that were almost equal to those using restored LJspeech.
Furthermore, in Common\,Voice, despite we trained a single-speaker TTS model using multiple speakers samples, the MOS of the TTS generated samples was 3.63.
We complement that we could not train the same WaveRNN neural vocoder using the original Common\,Voice dataset: the training failed to converge due to the noise in the target speech.
These results indicate that applying Miipher on a large scale speech dataset collected from the Web enables us to train high-quality TTS models.
Audio samples of the restored and TTS generated samples are available at our demo page\footref{footnote:demo}.

\section{Conclusions}
\label{sec:conclusion}

In this study, we proposed Miipher, a robust SR model based on the parametric re-synthesis framework for restoring speech samples in-the-wild.
Based on our preliminary investigation, we especially focused on two difficult degradation where SR frequently fails: phoneme masking and deletion.
To solve these issues, for the input features, we used speech and text representations extracted from w2v-BERT~\cite{w2vbert} and PnG-BERT~\cite{pngbert}, respectively.
Through experiments, we show that Miipher was robust against various audio degradation, could restore public speech samples from the Web, and enabled us to train high-quality TTS models form the restored samples.
From these results, we conclude that Miipher can increase the value of speech samples in-the-wild by improving the speech quality as the training data for speech generation tasks.

In the future, we will investigate the use of transcripts predicted using ASR to extract linguistic features and restore speech samples for which transcriptions do not exist. 
Also, we will develop multilingual Miipher in order to train high-quality TTS models for low-resource languages.
% The use of w2v-BERT trained on 16\,kHz sampling signals should not be suitable for Miipher: explicitly discarding information above 8\,kHz at the input of the feature-cleaner pushes full responsibility for generating the high-frequency content onto the neural vocoder. 
Training of SSL models to obtain speech representation from 24\,kHz sampling signals is also a future work.

\clearpage 
\balance
\bibliographystyle{IEEEtran}
\footnotesize
\bibliography{refs21}

\begin{thebibliography}{10}
\providecommand{\url}[1]{#1}
\def\UrlFont{\rmfamily}
\providecommand{\newblock}{\relax}
\providecommand{\bibinfo}[2]{#2}
\providecommand\BIBentrySTDinterwordspacing{\spaceskip=0pt\relax}
\providecommand\BIBentryALTinterwordstretchfactor{4}
\providecommand\BIBentryALTinterwordspacing{\spaceskip=\fontdimen2\font plus
\BIBentryALTinterwordstretchfactor\fontdimen3\font minus
  \fontdimen4\font\relax}
\providecommand\BIBforeignlanguage[2]{{%
\expandafter\ifx\csname l@#1\endcsname\relax
\typeout{** WARNING: IEEEtran.bst: No hyphenation pattern has been}%
\typeout{** loaded for the language `#1'. Using the pattern for}%
\typeout{** the default language instead.}%
\else
\language=\csname l@#1\endcsname
\fi
#2}}

\bibitem{Maiti_waspaa_2019}
S.~{Maiti} and M.~I. {Mandel}, ``Parametric resynthesis with neural vocoders,''
  in \emph{WASPAA}, 2019.

\bibitem{self_remaster}
T.~Saeki, S.~Takamichi, \emph{et~al.}, ``{SelfRemaster}: Self-supervised speech
  restoration with analysis-by-synthesis approach using channel modeling,'' in
  \emph{Interspeech}, 2022.

\bibitem{Su_2021}
J.~{Su}, Z.~{Jin}, and A.~{Finkelstein}, ``{HiFi-GAN-2}: Studio-quality speech
  enhancement via generative adversarial networks conditioned on acoustic
  features,'' in \emph{WASPAA}, 2021.

\bibitem{voice_filxer}
H.~{Liu}, Q.~{Kong}, \emph{et~al.}, ``{VoiceFixer}: Toward general speech
  restoration with neural vocoder,'' \emph{arXiv:2109.13731}, 2021.

\bibitem{UNIVERSE}
J.~Serrà, S.~Pascual, \emph{et~al.}, ``Universal speech enhancement with
  score-based diffusion,'' \emph{arXiv:2206.03065}, 2022.

\bibitem{dlwang_2018}
D.~Wang and J.~Chen, ``Supervised speech separation based on deep learning: An
  overview,'' \emph{IEEE/ACM TASLP}, vol.~26, no.~10, p. 1702–1726, 2018.

\bibitem{wpe}
T.~Nakatani, T.~Yoshioka, \emph{et~al.}, ``Speech dereverberation based on
  variance-normalized delayed linear prediction,'' \emph{IEEE TASLP}, 2010.

\bibitem{han_2014}
K.~Han, Y.~Wang, and D.~Wang, ``Learning spectral mapping for speech
  dereverberation,'' in \emph{ICASSP}, 2014.

\bibitem{dnn_wpe}
K.~Kinoshita, M.~Delcroix, \emph{et~al.}, ``Neural network-based spectrum
  estimation for online {WPE} dereverberation,'' in \emph{Interspeech}, 2017.

\bibitem{declip_survey}
P.~Záviška, P.~Rajmic, \emph{et~al.}, ``A survey and an extensive evaluation
  of popular audio declipping methods,'' \emph{IEEE J. Sel. Top. Signal
  Process.}, 2021.

\bibitem{Li_2015}
K.~Li and C.-H. Lee, ``A deep neural network approach to speech bandwidth
  expansion,'' in \emph{ICASSP}, 2015.

\bibitem{Kuleshov_2017}
V.~Kuleshov, S.~Z. Enam, and S.~Ermon, ``Audio super resolution using neural
  networks,'' in \emph{ICLR}, 2017.

\bibitem{wavenet}
A.~{van den Oord}, S.~{Dieleman}, \emph{et~al.}, ``{WaveNet}: A generative
  model for raw audio,'' \emph{arXiv:1609.03499}, 2016.

\bibitem{tacotron2}
J.~{Shen}, R.~{Pang}, \emph{et~al.}, ``Natural {TTS} synthesis by conditioning
  {WaveNet} on mel spectrogram predictions,'' in \emph{ICASSP}, 2018.

\bibitem{parrotron}
F.~Biadsy, R.~J. Weiss, \emph{et~al.}, ``{Parrotron}: An end-to-end
  speech-to-speech conversion model and its applications to hearing-impaired
  speech and speech separation,'' in \emph{Interspeech}, 2019.

\bibitem{speechstew}
W.~{Chan}, D.~{Park}, \emph{et~al.}, ``Speechstew: Simply mix all available
  speech recognition data to train one large neural network,''
  \emph{arXiv:2104.02133}, 2021.

\bibitem{bigvgan}
S.-g. Lee, W.~Ping, \emph{et~al.}, ``{BigVGAN}: A universal neural vocoder with
  large-scale training,'' in \emph{ICLR}, 2023.

\bibitem{wavefit}
Y.~Koizumi, K.~Yatabe, \emph{et~al.}, ``{WaveFit}: An iterative and
  non-autoregressive neural vocoder based on fixed-point iteration,'' in
  \emph{SLT}, 2023.

\bibitem{w2v22020}
A.~{Baevski}, H.~{Zhou}, \emph{et~al.}, ``{wav2vec 2.0}: A framework for
  self-supervised learning of speech representations,'' in \emph{NeurIPS},
  2020.

\bibitem{w2vbert}
Y.-A. Chung, Y.~Zhang, \emph{et~al.}, ``{w2v-BERT}: Combining contrastive
  learning and masked language modeling for self-supervised speech
  pre-training,'' in \emph{ASRU}, 2021.

\bibitem{bestrq}
C.-C. Chiu, J.~Qin, \emph{et~al.}, ``Self-supervised learning with
  random-projection quantizer for speech recognition,'' in \emph{ICML}, 2022.

\bibitem{common_voice}
R.~Ardila, M.~Branson, \emph{et~al.}, ``{Common Voice}: A
  massively-multilingual speech corpus,'' in \emph{LREC}, 2020.

\bibitem{vox_populi}
C.~Wang, M.~Riviere, \emph{et~al.}, ``{V}ox{P}opuli: A large-scale multilingual
  speech corpus for representation learning, semi-supervised learning and
  interpretation,'' in \emph{ACL}, 2021.

\bibitem{hsu2018hierarchical}
W.-N. Hsu, Y.~Zhang, \emph{et~al.}, ``Hierarchical generative modeling for
  controllable speech synthesis,'' in \emph{ICLR}, 2019.

\bibitem{hsu_2019}
------, ``Disentangling correlated speaker and noise for speech synthesis via
  data augmentation and adversarial factorization,'' in \emph{ICASSP}, 2019.

\bibitem{borsos22_interspeech}
Z.~Borsos, M.~Sharifi, and M.~Tagliasacchi, ``{SpeechPainter}: Text-conditioned
  speech inpainting,'' in \emph{Interspeech}, 2022.

\bibitem{pngbert}
Y.~Jia, H.~Zen, \emph{et~al.}, ``{PnG} {BERT}: Augmented {BERT} on phonemes and
  graphemes for neural {TTS},'' in \emph{Interspeech}, 2021.

\bibitem{hubert_neural_vocoder}
A.~Polyak, Y.~Adi, \emph{et~al.}, ``Speech resynthesis from discrete
  disentangled self-supervised representations,'' in \emph{Interspeech}, 2021.

\bibitem{film}
E.~Perez, F.~Strub, \emph{et~al.}, ``{FiLM}: Visual reasoning with a general
  conditioning layer,'' in \emph{AAAI}, 2018.

\bibitem{Koizumi_waspaa_2021}
Y.~Koizumi, S.~Karita, \emph{et~al.}, ``{DF-Conformer}: Integrated architecture
  of {Conv-TasNet} and {Conformer} using linear complexity self-attention for
  speech enhancement,'' in \emph{WASPAA}, 2021.

\bibitem{wavegrad}
N.~{Chen}, Y.~{Zhang}, \emph{et~al.}, ``{WaveGrad}: Estimating gradients for
  waveform generation,'' in \emph{ICLR}, 2021.

\bibitem{parallel_wavegan}
R.~Yamamoto, E.~Song, and J.-M. Kim, ``Parallel {WaveGAN}: A fast waveform
  generation model based on generative adversarial networks with
  multi-resolution spectrogram,'' in \emph{ICASSP}, 2020.

\bibitem{specgrad}
Y.~Koizumi, H.~Zen, \emph{et~al.}, ``{SpecGrad}: Diffusion probabilistic model
  based neural vocoder with adaptive noise spectral shaping,'' in
  \emph{Interspeech}, 2022.

\bibitem{Kong_2020}
J.~{Kong}, J.~{Kim}, and J.~Bae, ``{HiFi-GAN}: Generative adversarial networks
  for efficient and high fidelity speech synthesis,'' in \emph{NeurIPS}, 2020.

\bibitem{conformer}
A.~{Gulati}, C.-C. {Chiu}, \emph{et~al.}, ``Conformer: Convolution-augmented
  transformer for speech recognition,'' in \emph{Interspeech}, 2020.

\bibitem{quan2022_odyssey}
Q.~Wang, Y.~Yu, \emph{et~al.}, ``Attentive temporal pooling for
  {Conformer}-based streaming language identification in long-form speech,'' in
  \emph{Odyssey}, 2022.

\bibitem{dvector_data}
J.~Pelecanos, Q.~Wang, \emph{et~al.}, ``Parameter-free attentive scoring for
  speaker verification,'' in \emph{Odyssey}, 2022.

\bibitem{drvector}
J.~Pelecanos, Q.~Wang, and I.~L. Moreno, ``{Dr-Vectors}: Decision residual
  networks and an improved loss for speaker recognition,'' in
  \emph{Interspeech}, 2021.

\bibitem{tau_2021_dataset}
S.~{Wang}, A.~{Mesaros}, \emph{et~al.}, ``A curated dataset of urban scenes for
  audio-visual scene analysis,'' in \emph{ICASSP}, 2021.

\bibitem{image_method}
J.~B. {Allen} and D.~A. {Berkley}, ``Image method for efficiently simulating
  small-room acoustics,'' \emph{J. Acoust. Soc. Am.}, 1979.

\bibitem{demand}
J.~Thiemann, N.~Ito, and E.~Vincent, ``The diverse environments multi-channel
  acoustic noise database ({DEMAND}): A database of multichannel environmental
  noise recordings,'' \emph{Meet. Acoust.}, 2013.

\bibitem{yu_asr_2020}
Y.~Zhang, J.~Qin, \emph{et~al.}, ``Pushing the limits of semi-supervised
  learning for automatic speech recognition,'' in \emph{NeurIPS SAS 2020
  Workshop}, 2020.

\bibitem{NEURIPS2018_6832a7b2}
Y.~Jia, Y.~Zhang, \emph{et~al.}, ``Transfer learning from speaker verification
  to multispeaker text-to-speech synthesis,'' in \emph{NeurIPS}, 2018.

\bibitem{chen2018sample}
Y.~Chen, Y.~Assael, \emph{et~al.}, ``Sample efficient adaptive
  text-to-speech,'' in \emph{ICLR}, 2019.

\bibitem{nat}
J.~Shen, Y.~Jia, \emph{et~al.}, ``Non-attentive {Tacotron}: Robust and
  controllable neural {TTS} synthesis including unsupervised duration
  modeling,'' \emph{arXiv:2010.04301}, 2020.

\bibitem{wavernn}
N.~{Kalchbrenner}, W.~{Elsen}, \emph{et~al.}, ``Efficient neural audio
  synthesis,'' in \emph{ICML}, 2018.

\bibitem{libritts}
H.~Zen, R.~Clark, \emph{et~al.}, ``{LibriTTS}: A corpus derived from
  {LibriSpeech} for text-to-speech,'' in \emph{Interspeech}, 2019.

\bibitem{ljspeech17}
K.~Ito and L.~Johnson, ``The {LJ} speech dataset,''
  \url{https://keithito.com/LJ-Speech-Dataset/}, 2017.

\end{thebibliography}

\end{sloppy}
\end{document}